# Design Multimedia Expert Diagnosing Diseases System Using Fuzzy Logic (MEDDSFL)

Mohammed Salah Ibrahim[*a] and Doaa Waleed Al-Dulaimee [b]

[a] College of Computer, University of Anbar, City Anbar, Country Iraq
[b] College of Computer, University of Anbar, City Anbar, Country Iraq

*Abstract*

*In this paper we designed an efficient expert system to diagnose diseases for human beings. The system depended on several clinical features for different diseases which will be used as knowledge base for this system. We used fuzzy logic system which is one of the most expert systems techniques that used in building knowledge base of expert systems. Fuzzy logic will be used to inference the results of disease diagnosing. We also provided the system with multimedia such as videos, pictures and information for most of disease that have been achieved in our system. The system implemented using Matlab ToolBox and fifteen diseases were studied. Five cases for normal, affected and unaffected people's different diseases have been tested on this system. The results show that system was able to predict the status whether a human has a disease or not accurately. All system results are reported in tables and discussed in detail.*

ملخص (عربي)

في هذا البحث قمنا بتصميم نظام خبير لتشخيص الامراض البشرية. تم اعتماد عدة خصائص لعدة امراض واستخدامها من قبل النظام في استنتاج و تشخيص الامراض. استخدمنا النظام المنطقي المضبب والذي يعتبر واحد من اكثر تقنيات الانظمة الخبيرة المستخدمة في بناء الانظمة الخبيرة. يعمل النظام المنطقي المضبب على استنتاج وتشخيص الامراض. تم تزويد النظام بالصور والفيديو والمعلومات الإضافية لأغلب الامراض. تم تصميم وتنفيذ النظام باستخدام برنامج (*Matlab ToolBox*) حيث تم الاعتماد على خمسة عشر مرض وتم تنفيذ خمسة حالات لكل مرض على اشخاص مصابين واشخاص غير مصابين وتم توثيق النتائج في جداول.

## 1. Introduction

Decision systems often rely on historical information for the formulation of a best course of action. Storing, retrieving and checking the large volumes of data and information for consistency represents one of the main challenges in building decision support systems that use historical data[1]. Expert systems is one of decision systems that has wide using in the field of medical to solve several problems. Disease diagnosing is one of the most important problem in medical field. Human disease diagnosis is a complicated process and requires high level of expertise[2].

The expert systems can't cover all human disease in one systems so, we take 15 diseases with all its rules that system will needed in its knowledge base to be used in inference the degree of disease infection. The system will contain multimedia for most disease such as information, pictures and videos for every disease. the system will take input as some of the symptoms for certain disease and it will process this input and produce output that detect degree of infection in that disease. The work was implemented and reported in tables.

The rest of the paper is organized as follows. Related works are described in Section 2. Section 3 presents Fuzzy Expert Systems. Section 4 presents medical disease and clinical feature. Section 5 illustrate the Multimedia Expert Diagnosing Diseases System Using Fuzzy Logic. In section 6, experimental results are presented. Finally, some concluding remarks are presented in Section 7.

## 2. Related Works

There several literature reviews for using expert systems in the field of medical such as Nan-Chen Hsieh et. al. developed a clinical research information system (CRIS) for cardiovascular disease to facilitate surgery treatment tracking. The CRIS tracks hundreds of pieces of data through surgical stages and converts these data into computerized registries, provides surveillance mechanisms, and generates clinical interpretive reports in a timely manner. Surgeons can use the CRIS to identify surgical-related data and interventional cardiovascular procedure risks based on specific patient characteristics, and it has increased the quality and efficiency of patient care [3].

Hui-Ling Chen et. al. presented an effective and efficient diagnosis system using fuzzy k-nearest neighbor (FKNN) for Parkinson's disease (PD) diagnosis. The proposed FKNN-based system is compared with the support vector machines (SVM) based approaches. In order to further improve the diagnosis accuracy for detection of PD, the principle component analysis was employed to construct the most discriminative new feature sets on which the optimal FKNN model was constructed [4].

Mahdi Jampour et. al. used the fuzzy logic model approach to determine and calculate lack or involvement of each the possible disease with neurological signs and sufficiently reduced natural Uncertainty regarding the diagnosis of disease [5].

Mir Anamul Hasan et. al. described a project work aiming to develop a web-based fuzzy expert system for diagnosing human diseases. Now a days fuzzy systems are being used successfully in an increasing number of application areas; they use linguistic rules to describe systems. This research project focuses on the research and development of a web-based clinical tool designed to improve the quality of the exchange of health information between health care professionals and patients [2].

Putu Manik Prihatini and I Ketut Gede Darma Putra developed an expert system combines the method of Fuzzy Logic and Certainty Factors with the object of research is a disease of tropical infectious diseases include Dengue Fever, Typhoid Fever and Chikungunya. Fuzzy logic methods will be used to handle the uncertainty experienced the patient"s symptoms and the certainty factor method will be used to handle the inability of an expert in defining the relationship between the symptoms of the disease with certainty. Expert system developed on web based platform, provide improve of knowledge, where expert can add new knowledge to a disease or alter the existing knowledge on the disease, so the system will remain accurate and up to date[6].

Sanjeev Kumar and Gursimranjeet Kaur presented system to detect the heart diseases in the person by using Fuzzy Expert System. The designed system based on the Parvati Devi hospital, Ranjit Avenue and EMC hospital Amritsar and International Lab data base. The system consists of 6 input fields and two output field. Input fields are chest pain type, cholesterol, maximum heart rate, blood pressure, blood sugar, old peak. The output field detects the presence of heart disease in the patient and precautions accordingly [7].

Ali.Adeli, Mehdi.Neshat presented a Fuzzy Expert System for heart disease diagnosis. The designed system based on the V.A. Medical Center, Long Beach and Cleveland Clinic Foundation data base. The system has 13 input fields and one output field. Input fields are chest pain type, blood pressure, cholesterol, resting blood sugar, maximum heart rate, resting electrocardiography (ECG), exercise, old peak (ST depression induced by exercise relative to rest), thallium scan, sex and age. The output field refers to the presence of heart disease in the patient [8].

Malathi A. and Santra A. K used neural networks and fuzzy logic in the medical field (carcinogenesis (pre-clinical study)). In carcinogenesis, neuro-fuzzy have been successfully applied to the problems in both pre-clinical and post-clinical diagnosis [9].

Ersin Kaya et.al. used fuzzy rule-based classifier for the diagnosis of congenital heart disease. Congenital heart diseases are defined as structural or functional heart disease. Medical data sets were obtained from Pediatric Cardiology Department at Selcuk University, from years 2000 to 2003. Firstly, fuzzy rules were generated by using medical data. Then the weights of fuzzy rules were calculated [10].

Our system will use the fuzzy logic system with mamdani inference model and centroid defuzzification method. Our system named as Multimedia Expert Diagnosing Diseases System Using Fuzzy Logic (MEDDSFL) which included with multimedia to increase the interactivity with the users.

## 3. Fuzzy Expert Systems

A computer Program Capable of performing at a human- expert level in a narrow problem domain area is called an expert system [11]. An expert system that uses fuzzy logic instead of Boolean logic is known as Fuzzy expert system. A fuzzy expert system is a collection of fuzzy rules and membership functions that are used to reason about data. Using fuzzy expert system expert knowledge can be represented that use vague and ambiguous terms in computer Fuzzy [2].

Fuzzy logic originates also from artificial intelligence field, it was developed by L. Zadeh in the seventies of the last century [12].

Fuzzy logic is a set of mathematical principles for knowledge representation based on degrees of membership rather than the crisp membership of classical binary logic. Unlike two-valued Boolean logic, fuzzy logic is multi valued. Fuzzy logic is a logic that describes fuzziness. As fuzzy logic attempts to model human's sense of words, decision making and common sense, it is leading to more human intelligent machines [13].

FL incorporates a simple, rule-based IF X AND Y THEN Z approach to a solving control problem rather than attempting to model a system mathematically. A FS consists of three main steps: Fuzzification, Rules base and Inference engine, and Defuzzification. These steps and the general architecture of a FS are shown in Figure 2 [14].

This template provides authors with most of the formatting specifications needed for preparing electronic versions of their papers. All standard paper components have been specified for three reasons: (1) ease of use when formatting individual papers, (2) automatic compliance to electronic requirements that facilitate the concurrent or later production of electronic products, and (3) conformity of style throughout a conference proceedings. Margins, column widths, line spacing, and type styles are built-in; examples of the type styles are provided throughout this document and are identified in italic type, within parentheses, following the example. PLEASE DO NOT RE-ADJUST THESE MARGINS. Some components, such as multi-leveled equations, graphics, and tables are not prescribed, although the various table text styles are provided. The formatter will need to create these components, incorporating the applicable criteria that follow.

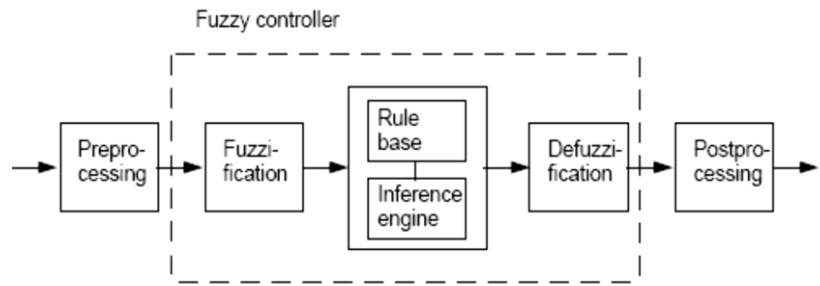

Fig.1 Fuzzy Logic Steps

The three steps are explained as follow [15]:
2. Fuzzification: in this step we will define the linguistic verbs of the problem and determine the range of these terms to draw the membership function for each term.
3. Inference engine: in this step we will firstly build the rule base that will use by the inference engine to determine the output. There is different number of inference engine but the most used are The Mamdani (which use max and min fuzzy set operations) and Takagi-Sugeno (which use equations).
4. Finally, defuzzification: in this step the fuzzy output will be converted to the crisp output.

**4. Medical Diseases and Clinical Feature**

The record of human suffering and death caused by smallpox, cholera, typhus, dysentery, malaria, etc. establishes the eminence of the infectious diseases. Despite the outstanding successes in control afforded by improved sanitation, immunization, and antimicrobial therapy, the infectious diseases continue to be a common and significant problem of modern medicine. The most common disease of mankind, the common cold, is an infectious disease, as is the feared modern disease AIDS. Some chronic neurological diseases that were thought formerly to be degenerative diseases have proven to be infectious. There is little doubt that the future will continue to reveal the infectious diseases as major medical problems [16].

Most medical diseases are detected depending on their clinical features that have been discover from the trying and failed of experience doctors. Table 1 illustrate some clinical features for some diseases that have been used in our system.

Table 1 Diseases and its Clinical Features

| No. | Disease | Clinical Features |
|---|---|---|
| 1 | Gout | A. recurent attacks of articular and periatricular inflammation.<br>B. tophi,crystals deposition in[articular , osseous,soft tissues].<br>C. uric acid in the urinary track.<br>D. interstitial nephropathy with renal function impairment. |
| 2 | Malaria | A. jaundice is commen dueto hemolytic hepatic destruction.<br>B. enlargement of liver & spleen with tenderness.<br>C. anemia developed rapidly.<br>D. patient may develop serious complications<br>E. children may rapidly die<br>F. relapse up to l year. |
| 3 | Typhoid | A. the patient develop fever, headache, malaise, anorexia, nausea, abdominal pain, myalgias, arthralgia, cough and sore throat. |
| 4 | Jaundice | A. yellowish discoloration of skin and sclera. |

|   |   | Indirect | 1.organish-yellow | 2.stool:normal color | 3.urine:normal |
|---|---|---|---|---|---|
|   |   | Direct | 1.greenish-yellow | 2.stool:clay(white) | 3.tea color |
| 5 | Diabetes Mellitus | A. polyuria<br>B. polydipsia<br>C. thirsty.<br>D. loss of weight. | | | |
| 6 | Shock | A. tachycardia at begening then shifting to bradycardia.<br>B. hypotension.<br>C. sweating.<br>D. cold extremidies.<br>E. loss of conscious. | | | |
| 7 | Hepatitis | A. nausea.<br>B. fever.<br>C. fatigue.<br>D. abdomina discomfort.<br>E. jaundice.<br>F. hepatomegaly. | | | |

## 5. Design of MEDDSFL System

The MEDDSFL model implemented using Matlab Toolbox which is a high-performance language for technical computing. In MEDDSFL system, we try to deal with a non-expert user and benefit from simple and clever ideas to diagnose case of person. If the doctor or even the person did not succeed in evaluating the case of person, the application gives the user of the system advice on what he/she should do. The Graphical User Interface (GUI) performs as a communication tool that connects the user with the system. It displays the inputs that entered by the user about the clinical feature that he feel it to be answered by the system and shows the corresponding results .

Before discussion results of implementing the MEDDSFL , we introduce the relationships between inputs and outputs by using the surface viewer which is a tool provided by Matlab. This tool displays the relationships between inputs and outputs in 3-diminsional surfaces. For example, we take typhoid disease as one of the diseases that we diagnosing by fuzzy system. The relationships between temperature of body, nausea and headache with the disease output are illustrated in figures (2, 3, 4 and 5) respectively. Let's take example to understand the relationship between the inputs and outputs, in figure 3 we saw that the slop of disease output is down when the temperature is 40 degree and nausea is 60 and the slop is begin toward high when the temperature and nausea are high. Other relationships have the same concept. And so on for each disease that we covered in MEDDSFL system such as jaundice, dehydration, pneumonia, nephrotic syndrome, thyrotoxia, diabetes, heart failure, and other diseases.

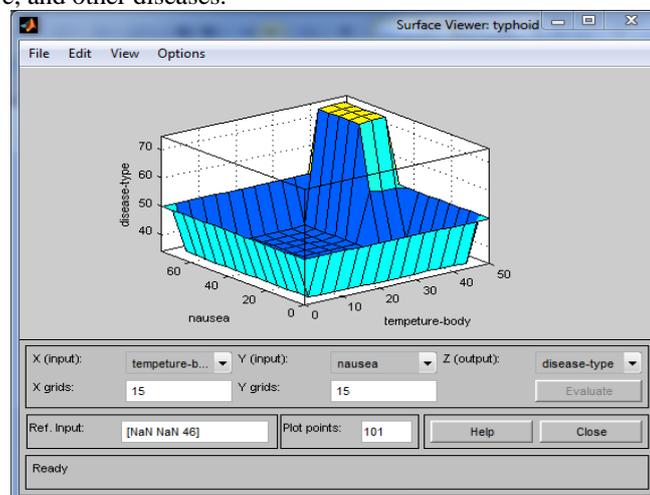

Fig. 2 Show 3d relationship between nausea, temperature-body, and disease of the person

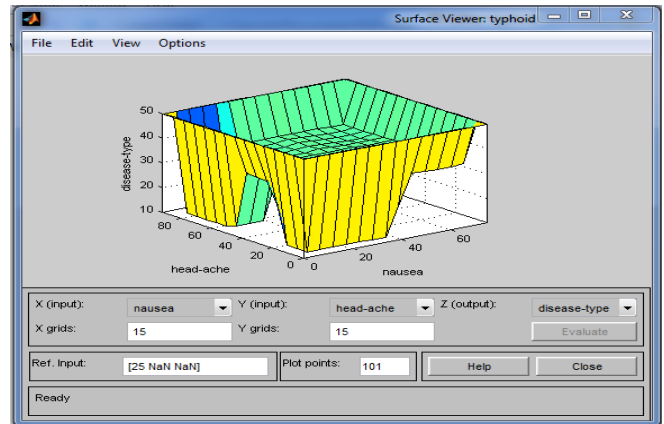
Fig. 3 show 3d relationship between nausea, headache and disease of the person.

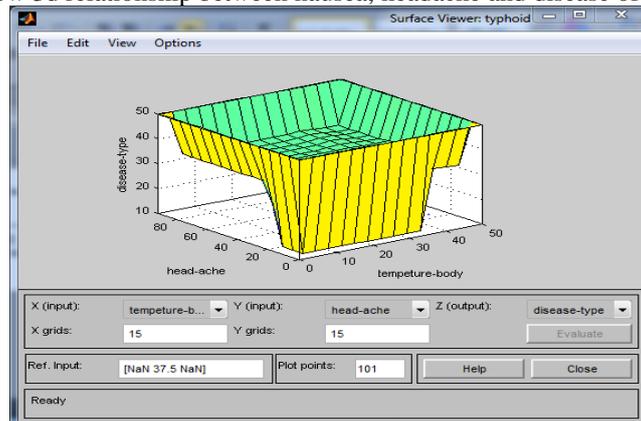
Fig. 4 show 3d relationship between temperature ,head-ache and disease of the person.

The system has interactive window (see fig. 5). choosing the disease from the pop-up menu that we included all the diseases that we refer to it Previously. When click on the elements of the menu, the window of your choice disease will be appear.

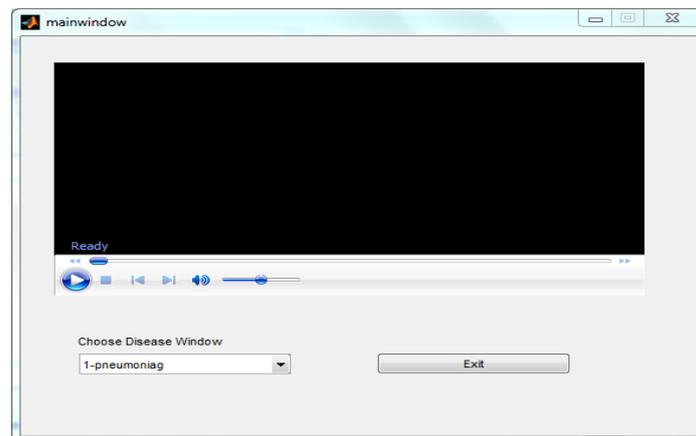
Fig. 5 show the main window in MEDDSFL system.

*5.1 Explanation for MEDDSFL Interface*

Figure 6 illustrate the interface of one of the MEDDSFL disease system which is shock disease. The inputs of the system represent one of the clinical features and the user represents the doctor, person or medical student who want to use the system to get information about specific disease, he must enter the clinical features in the inputs group textboxes after that he click the execution button to get the suitable output. The MEDDSFL provide two type of output, one as crisp value and another is the linguistic variable that equivalent to the crisp output.

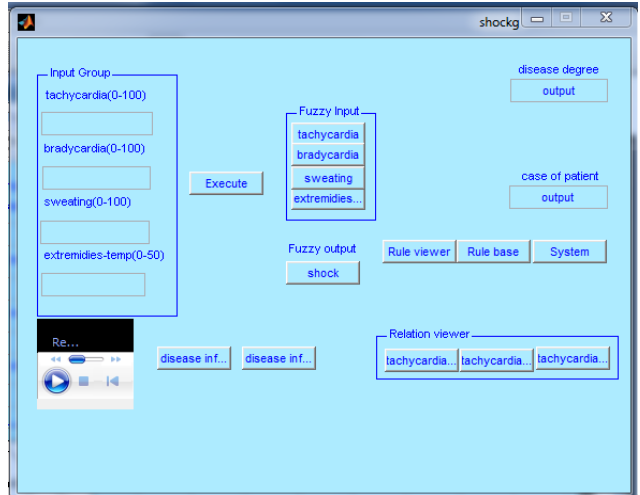
Fig 6 GUI for shock disease before linking

The MEDDSFL disease interface provide fast access to the diseases windows and also you can see the general form of the selected disease by clicking the "System" button and also you can see the Rule Base of the system, Rule Viewer and you can track the relationships between the inputs and outputs by clicking on the their button and you can see multimedia about the disease such as read some of information that we supported the system with it by clicking the "DISEASES INFORMATION" and you can see videos speaking about the disease by clicking to small video window that we added to the GUI interface by using Active-x control.

## 6. Experimental Results

The system implemented using Matlab ToolBox and fifteen disease were achieved for difference diseases. five cases for normal, affected and unaffected persons for fifteen diseases have been tested on our system. The results have been reported in Table 2 . The inputs (clinical feature) are taken in different forms such as person has tachycardia 67.9 , bradycardia 75.8, sweating 34.3, and extremidies-temp 28 , any experience one will take in mind these variable and begin by inferenecing , in this case he will say: this person is need analysis to ensure that if the person injected or not with the shock disease see fig. 7.

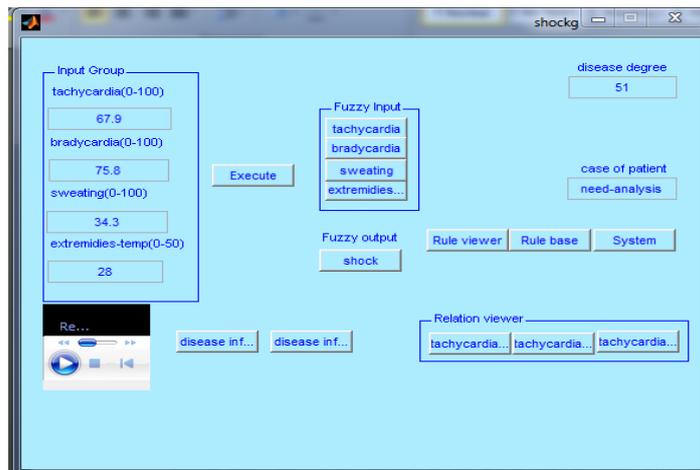
Fig. 7 Case Of Shock Inference Result

Table 2 Results of implementing MEDDSFL system with 5 symptoms for each disease

| NO. | DISEASE TYPE | FIRST INPUT | SECOND INPUT | THIRD INPUT | FOURTH INPUT | INFERENCE OUTPUT |
|---|---|---|---|---|---|---|
| 1 | Dehydration | Sunken eyes | Drowsy | Tongue | Extremities | |
| | | 84.1 | 80.8 | 65.9 | 34.6 | Need analysis(47) |
| | | 70 | 60 | 50 | 50 | Injected (69) |
| | | 73.48 | 79.23 | 65.91 | 90 | Injected (79) |
| | | 23.48 | 16.15 | 12.88 | 11.54 | not-injected (18) |
| | | 41.67 | 13.08 | 31.06 | 45.38 | not-injected (20) |
| 2 | Diabetes | Polyuria | Polydipsia | Loss of weight | | |
| | | 25.55 | 5 | 25 | | Not injected (38) |
| | | 15.36 | 1.265 | 14.16 | | Not-injected (15) |

| # | Disease | | | | | |
|---|---|---|---|---|---|---|
| | | 15.36 | 14.16 | 40.06 | | Need-analysis (50) |
| | | 33.43 | 7.53 | 40.06 | | Injected (62) |
| | | 6.928 | 5.241 | 8.133 | | Not-injected (34) |
| 3 | Duodenal ulcer | Epigastria distress | Vomiting | Heart burn | | |
| | | 8.6 | 82.5 | 78.9 | | Injected (83.2) |
| | | 76.51 | 87.35 | 50 | | Need-analysis(64) |
| | | 94.58 | 83.73 | 99.4 | | Injected (83.2) |
| | | 30.72 | 23.49 | 50 | | Not-injected(35) |
| | | 77.7 | 31.93 | 18.67 | | injected(83) |
| 4 | Heart failure | Orthopnea | Crepitation | Cyanosis | | |
| | | 83.7 | 84.9 | 68.1 | | Not injected (20.1) |
| | | 64.46 | 50 | 82.53 | | Not injected (31.5) |
| | | 21.08 | 21.08 | 16.27 | | Injected (78.8) |
| | | 70.48 | 60.84 | 68.07 | | Not injected (19.7) |
| | | 30.72 | 25.9 | 56.02 | | Need-analysis (51) |
| 5 | Hepatitis | Nausea | temperature of body | Jaundice | | |
| | | 37.5 | 25.22 | 37.5 | | Not injected (30.5) |
| | | 16.72 | 37.65 | 56.48 | | Not injected (29) |
| | | 65.51 | 40.66 | 56.48 | | Need-analysis (50) |
| | | 66.42 | 47.29 | 67.32 | | Injected (65.6) |
| | | 42.02 | 47.29 | 67.32 | | Not injected (49) |
| 6 | hypoglycemia | Apnea | Pallor | Seizure | Temperature of body | |
| | | 88.6 | 83.5 | 88.6 | 47.3 | Need analysis (49) |
| | | 38.64 | 3.846 | 78.03 | 20.38 | Not injected (18.1) |
| | | 65.91 | 34.62 | 71.97 | 45 | Need analysis (50) |
| | | 15.91 | 50 | 56.82 | 19.62 | Not injected (18) |
| | | 47.73 | 53.08 | 94.7 | 37.31 | Not injected (33.2) |
| 7 | jaundice | Skin color | Stool | Urine | | |
| | | 8.37 | 8.61 | 8.49 | | Need analysis (8) |
| | | 4.398 | 6.928 | 8.855 | | Need analysis (6) |
| | | 3.675 | 3.675 | 0.6627 | | Not-injected (2.02) |
| | | 8.614 | 8.735 | 8.735 | | injected (8.53) |
| | | 5.361 | 5 | 4.639 | | Not-injected (2) |
| 8 | malaria | Jaundice | Hemolysis | Anemia | | |
| | | 59.6 | 19.4 | 6.7 | | Need analysis (54) |
| | | 93.37 | 70.03 | 7.663 | | Injected (70) |
| | | 93.37 | 48.34 | 4.675 | | Need analysis (61) |
| | | 59.64 | 24.85 | 4.675 | | Need analysis (54) |
| | | 40.36 | 29.37 | 6.892 | | Not-Injected (49) |
| 9 | Meningitis encephala | Temperature of body | Head-ache | Loss of consciousness | Nick stiffness | |
| | | 39.6 | 82.9 | 70.3 | 56.3 | Injected (89.4) |
| | | 39.6 | 20 | 34 | 45 | need-analysis (61) |
| | | 40 | 50 | 82 | 45 | need-analysis (73) |
| | | 25 | 77.69 | 81.06 | 50 | Injected (88.4) |
| | | 1.894 | 11.54 | 21.97 | 33.08 | Not-injected (24) |
| 10 | Nephrotic syndrome | Swelling legs and feet | Loss of appetite | Diarrhea | | |
| | | 40.4 | 35.5 | 63.3 | | Need analysis (49) |
| | | 12.65 | 54.82 | 71.69 | | Not-injected (26) |
| | | 80.12 | 88.55 | 31.93 | | Need analysis (49) |
| | | 80.12 | 43.98 | 71.69 | | Injected (82.2) |
| | | 96.99 | 53.61 | 39.16 | | need-analysis (61) |
| 11 | pneumonia | illness of cough | Temperature of body | Chest pain | Shortness of | |
| | | 53.8 | 36.5 | 70.5 | 66.9 | Need analysis (56) |
| | | 90.15 | 43.46 | 90.15 | 88.46 | Injected (83) |
| | | 55.3 | 21.15 | 18.94 | 43.85 | Need analysis (56) |
| | | 11.36 | 37.31 | 11.36 | 43.85 | Not-injected (22) |
| | | 50.76 | 36.54 | 38.64 | 25.38 | Not-injected (31) |
| 12 | Rickets | Curvature of spine | Skeletal curvature of legs | Weakness of tooth enamel | | |
| | | 77.4 | 68.2 | 61.6 | | Injected (81) |
| | | 87.35 | 71.69 | 94.58 | | Injected (80) |
| | | 34.34 | 28.31 | 52.41 | | Not-injected(29) |
| | | 46.39 | 56.02 | 51.2 | | Need analysis (56) |
| | | 74.1 | 25.9 | 64.46 | | Need analysis (50) |

| | | Tachyc-ardia | Bradycardia | Sweating | Extreminidies temperature | |
|---|---|---|---|---|---|---|
| **13** | shock | 76.5 | 26.9 | 46.2 | 39.2 | Need analysis (51) |
| | | 50 | 43.85 | 50 | 37.69 | Not-injected (29) |
| | | 91.67 | 90 | 14.39 | 37.69 | Need analysis (51) |
| | | 91.67 | 80.77 | 81.06 | 88.46 | Injected (81) |
| | | 31.06 | 33.08 | 14.39 | 0.7692 | Not-injected (20) |
| | | Weight loss | Palpitation | Tachycardia | | |
| **14** | thyrotoxicosis | 84.9 | 36.7 | 60.8 | | Need analysis (54) |
| | | 50 | 50 | 53.61 | | Need-analysis (62) |
| | | 58.43 | 80.12 | 12.65 | | Not-injected (80) |
| | | 48.8 | 52.41 | 46.39 | | Need-analysis (68) |
| | | 48.8 | 71.69 | 45.18 | | Injected (73.2) |
| | | Temperature of body | Nausea | Head-ache | | |
| **15** | typhoid | 31.6 | 18.5 | 23.8 | | Not injected (10) |
| | | 40.66 | 49.25 | 75.93 | | Injected (2) |
| | | 16.57 | 32 | 26.05 | | Not-injected (23.6) |
| | | 44.28 | 57.38 | 80.36 | | Injected (76) |
| | | 37.05 | 20. 33 | 41.57 | | Not-injected(35) |

**7. Conclusions**

In this project, a Knowledge-Based System (KBS) for Multimedia Expert Diseases Diagnosing System via Fuzzy logic (MEDDSFL) is presented. The MEDDSFL is proposed to assist doctors in diseases diagnosing and medicine students to automatically diagnose diseases without the help of human expert. In addition, MEDDSFL considered as an interactive training tool that can provide expert guidance in diseases diagnosing using Fuzzy logic system. System results indicate that the fuzzy logic system was able to provide accurate prediction ratios and showed the ability to have a medical diagnosing system that can act as a doctor. For future work, Further improvement to the system domain knowledge base is required to enhance system predictability. Furthermore, adopting other AI techniques and another disease's parameters to cover more diseases and get more accurate results.

[16] Samuel Baron, "Medical Microbiology", 4th edition, Galveston (TX), 1996.

**Mohammed Salah Ibrahim Al-Obaidi** Obtained B.Sc. (2008), M.Sc. (2011) in field of Computer Science from the College of Computer, University of Anbar. Worked as an administrator in the Department of calculating in Al-Safa Co. to oversee the reconstruction works in Iraq (2007), and worked with Afaq Co. as accountant (2010). Now, Mohammed is Faculty staff member in Computer Science Department in College of Computer, University of Anbar. His current 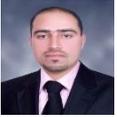 interesting field focuses on Cloud Computing, Wireless Network Management, Artificial Intelligence and Network Security.